\documentclass[aps,prd,twocolumn,10pt,groupedaddress]{revtex4-1}
\usepackage{amssymb}
\usepackage{graphicx}
\usepackage{amsmath}
\usepackage{hyperref}
\usepackage{subfigure}
\usepackage{float}
\usepackage{color}
\usepackage{ulem}
\usepackage[utf8]{inputenc}

\begin{document}

\title{A refined trans-Planckian censorship conjecture}
\author{Rong-Gen Cai}
\email{cairg@itp.ac.cn}
\affiliation{CAS Key Laboratory of Theoretical Physics, Institute of Theoretical Physics, Chinese Academy of Sciences, Beijing 100190, China}
\author{Shao-Jiang Wang}
\email{schwang@cosmos.phy.tufts.edu}
\affiliation{Tufts Institute of Cosmology, Department of Physics and Astronomy, Tufts University, 574 Boston Avenue, Medford, Massachusetts 02155, USA}

\begin{abstract}
We propose a refined version of trans-Planckian censorship conjecture (TCC), which could be elaborated from the strong scalar weak gravity conjecture combined with some entropy bounds. In particular, no fine-tuning on the inflation model-building is required in the refined TCC, and it automatically passes the tests from those stringy examples that support the original TCC. Furthermore, our refined TCC could be consistent with hilltop eternal inflation.
\end{abstract}
\maketitle

\section{Introduction}\label{sec:int}

Decades of explorations along string theory as a promising candidate for quantum gravity reveal us a landscape \cite{Susskind:2003kw} of a huge number of low-energy effective theories, beyond which are conjectured mostly in the swampland \cite{Vafa:2005ui} that admit no consistent completion into quantum gravity in the ultraviolet (UV). The string-inspired swampland criteria (see, for example, a review \cite{Palti:2019pca}) have been proposed to discriminate those seemingly innocent theories consistent at low-energy in the infrared (IR). 

The difficulties of locating de Sitter (dS) vacua in string theory have inspired the swampland dS conjecture (SdSC) \cite{Obied:2018sgi} and its refinements \cite{Dvali:2018fqu,Andriot:2018wzk,Rudelius:2019cfh,Ooguri:2018wrx,Garg:2018reu,Andriot:2018mav}. The original SdSC \cite{Obied:2018sgi} simply forbids any dS local extrema, which is in direct tension with the standard model Higgs potential \cite{Denef:2018etk,Cicoli:2018kdo,Murayama:2018lie,Choi:2018rze,Hamaguchi:2018vtv} (see also \cite{Han:2018yrk}) that has a dS local maximum. Therefore, the refined SdSC \cite{Dvali:2018fqu,Andriot:2018wzk,Rudelius:2019cfh,Ooguri:2018wrx,Garg:2018zdg,Andriot:2018mav} was proposed to relax the constraint so that those local maxima with large second-derivative in field space are allowed, while the rest extrema are forbidden including all dS local minima. In particular, both requirements of no quantum broken dS \cite{Dvali:2018fqu,Dvali:2018jhn} and no eternal inflation \cite{Rudelius:2019cfh} disfavour any stable dS local minima. Similarly, the trans-Planckian censorship conjecture (TCC) \cite{Bedroya:2019snp} was recently proposed to further relax the constraint that, all modes that exit the Hubble horizon at the end of inflation should have their physical length larger than the Planck length at the beginning of inflation, namely, $a_i/(a_fH_f)>1/M_\mathrm{Pl}$. This strongly constrains the duration of inflation on the elapsed e-folding number,
\begin{align}\label{eq:TCC}
\mathrm{e}^N<\frac{M_\mathrm{Pl}}{H_f}.
\end{align}
Therefore, TCC further relaxes the swampland dS criteria by forbidding any stable dS local minima but sparing the life of metastable dS local minima. As an immediate result, inflation cannot be eternal both into the past \cite{Borde:2001nh} and future directions \cite{Dvali:2018fqu,Dvali:2018jhn,Rudelius:2019cfh,Seo:2019wsh}.

When applied to the inflationary phenomenology \cite{Bedroya:2019tba}, TCC sets an upper bound on the inflationary Hubble scale and corresponding tensor-to-scalar ratio by
\begin{align}
H&<\mathrm{e}^{-N}M_\mathrm{Pl}\approx\mathcal{O}(10)\,\mathrm{MeV},\\
r&\equiv\frac{2}{\pi^2\mathcal{P}_\mathcal{R}}\left(\frac{H}{M_\mathrm{Pl}}\right)^2<6.8\times10^{-33},
\end{align}
where $N=(1/3)\ln(M_\mathrm{Pl}/H_0)\approx46.2$ for an instantaneous reheating history \cite{Bedroya:2019tba,Mizuno:2019bxy} and $\mathcal{P}_\mathcal{R}\approx2.1\times10^{-9}$ from Planck 2018 \cite{Aghanim:2018eyx}. This imposes a severe fine-tuning problem on the inflationary model-building with initial condition on slow-roll parameter down to $\epsilon\approx10^{-33}$, which could be relaxed by invoking non-instantaneous reheating history \cite{Mizuno:2019bxy} at the price of ultra-low reheating temperature (see \cite{Cai:2019igo} for its implicaitons on the mass of primordial black holes). Other trials of model-building within TCC include inflationary dark matter \cite{Tenkanen:2019wsd}, warm inflation \cite{Das:2019hto,Goswami:2019ehb}, initial states \cite{Cai:2019hge,Brahma:2019unn}, non-standard post-inflationary history \cite{Dhuria:2019oyf,Torabian:2019zms}, D-term hybrid inflation \cite{Schmitz:2019uti}, multi-stage inflation \cite{Berera:2019zdd,Li:2019ipk,Torabian:2019qgl}, inflection-point inflation \cite{Okada:2019yne}, negative running of spectral index \cite{Kadota:2019dol}. Beyond the original statement of TCC, \cite{Lin:2019pmj} proposed a generalized TCC to save $k$-inflation \cite{ArmendarizPicon:1999rj} from the problem that sub-Planckian fluctuations could exit Hubble horizon without violating the original TCC. Furthermore, \cite{Saito:2019tkc} casts the doubt on TCC as an universal swampland criterion but proposes a probabilistic interpretation.

Although TCC requires all trans-Planckian quantum modes to remain their quantum nature by never exiting the Hubble horizon, the apparent physical consequences of having trans-Planckian fluctuations stretched out of the Hubble horizon are unclear (see, however, \cite{Dvali:2010bf,Dvali:2010jz,Dvali:2014ila} for self-completeness of Einstein gravity), which should be elaborated with other more established arguments. One inspiration comes from some earlier variants of refined SdSC \cite{Dvali:2018fqu,Dvali:2018jhn,Rudelius:2019cfh} that also prohibit long-lasting dS vacua. Therefore, although the refined SdSC in \cite{Ooguri:2018wrx} is silent about the dS lifetime, the arguments they used to obtain the refined SdSC could shed light on the physical interpretation of TCC. Recall that the refined SdSC \cite{Ooguri:2018wrx} was found following from the swampland distance conjecture (SDC) \cite{Ooguri:2006in} combined with the Bousso entropy bound \cite{Bousso:1999xy} applied to an accelerating universe, which could also be used in \cite{Seo:2019wsh} to derive an entropic quasi-dS instability time. In particular, TCC was recently elaborated from SDC \cite{Brahma:2019vpl} in the large field excursion limit, which could also be derived from some entropy bounds \cite{Kehagias:2019iem}. To elaborate TCC over the whole field space, we use a strong version \cite{Gonzalo:2019gjp} of scalar weak gravity conjecture (WGC) \cite{Palti:2017elp} combined with some entropy bounds, and then a weaker version of TCC is obtained to naturally evade the fine-tuning problem of initial conditions without turning to any exotic model-building.

The outline for this paper is as follows: In Sec. \ref{sec:SWGC}, we review the scalar WGC and then derive a bound for the species number from both scalar WGC and its strong version. In Sec. \ref{sec:TCC}, using previous bound on species number, we propose a refined TCC from two separate arguments of UV-IR hierarchy and entropy bounds. In Sec. \ref{sec:eternal}, the refined TCC is reformulated for eternal inflation. In Sec. \ref{sec:con}, we conclude the result and discuss the refined TCC implications for inflation without fine-tuning problem.

\section{Scalar weak gravity conjecture}\label{sec:SWGC}

The scalar WGC \cite{Palti:2017elp} implements the belief that the net force mediated by scalar fields should be larger than the gravitational force. In specific, one considers a complex scalar field $\varphi$ with a field-dependent mass term $m^2(\phi)|\varphi|^2$, where a canonically normalized scalar field $\phi$ mediates the scalar force between $\varphi$ particles via a trilinear coupling term $\partial_\phi m^2(\phi_0)\delta\phi|\varphi|^2$ when expanding $\phi=\phi_0+\delta\phi$ around a vacuum expectation value (vev) $\phi_0$. By requiring the scalar force larger than the gravity,
\begin{align}
F_\mathrm{scalar}\equiv\frac{(\partial_\phi m)^2}{4\pi r^2}>\frac{m^2}{8\pi M_\mathrm{Pl}^2r^2}\equiv F_\mathrm{gravity},
\end{align}
the scalar WGC reads
\begin{align}\label{eq:SWGC}
2(\partial_\phi m)^2>\frac{m^2}{M_\mathrm{Pl}^2},
\end{align}
Similar form also holds for fermion or massless multi-scalar fields $\phi^i$ with kinetic term $g_{ij}\partial\phi^i\partial\phi^j$, that is, there should have a state with mass $m$ satisfying the bound
\begin{align}
2g^{ij}(\partial_{\phi^i}m)(\partial_{\phi^j}m)>\frac{m^2}{M_\mathrm{Pl}^2}.
\end{align}

In particular, as an explicit fulfillment of \eqref{eq:SWGC}, SDC \cite{Ooguri:2006in} allows for a large distance replacement (corresponding to weak coupling limit) over the moduli space without a  potential (or field space in the effective theory with a potential \cite{Klaewer:2016kiy,Baume:2016psm}) if a infinite tower of states becomes light with mass scale
\begin{align}\label{eq:SDC}
m(\phi)\sim m(\phi_0)\mathrm{e}^{-\alpha\frac{|\Delta\phi|}{M_\mathrm{Pl}}}
\end{align}
descending from the UV with some $\mathcal{O}(1)$ constant $\alpha>0$. Note that the scalar WGC is formulated for any field value in field space, while the SDC holds most straightforwardly at large distances in field space, $|\Delta\phi|/M_\mathrm{Pl}\gtrsim1$. Therefore, our use of scalar WGC in replacement of SDC is a non-trivial generalization of the argument used in \cite{Seo:2019wsh}. Nevertheless, we all arrive at the same dS lifetime as shown in the end. 

\subsection{A bound for species number}\label{subsec:SWGC}

Since both scalar WGC and SDC could constrain the mass scale of an infinite tower of states for a scalar theory coupled to gravity, the effective description of the scalar theory might be jeopardized by including sufficient number of states. Therefore, there exit a bound on the number of states included, above which we are not in any weakly-coupled gravitational regime but a strong coupling regime. Or equivalently, for given $N_s$ particle states, gravity becomes strongly coupled above a dubbed species scale (conjecture) \cite{Veneziano:2001ah,Dvali:2001gx,ArkaniHamed:2005yv,Distler:2005hi,Dimopoulos:2005ac,Dvali:2007hz,Dvali:2007wp,Dvali:2009ks},
\begin{align}
\Lambda_s\equiv\frac{M_\mathrm{Pl}}{\sqrt{N_s}}.
\end{align}
Recall that, for example, the number of Kaluza-Klein (KK) modes $N_s$ we can include in the lower $d$ dimensional gravity theory before the true quantum gravity cutoff scale $M_\mathrm{Pl}^D$ of the higher $D$ dimensional gravity theory is $N_s\sim M_\mathrm{Pl}^D/m_\mathrm{KK}$. Likewise,  $N_s\sim\Lambda_s/m$ is expected in general. Hence the species number reads
\begin{align}\label{eq:Ns}
N_s\sim\left(\frac{M_\mathrm{Pl}}{m(\phi)}\right)^\frac23,
\end{align}
which, after the use of scalar WGC \eqref{eq:SWGC}, becomes
\begin{align}
(\partial_\phi\ln N_s)^2\gtrsim\frac29M_\mathrm{Pl}^{-2}.
\end{align}
This could be integrated directly to give
\begin{align}\label{eq:NsDphi}
\ln N_s\gtrsim\frac{\sqrt{2}}{3}\frac{|\Delta\phi|}{M_\mathrm{Pl}}.
\end{align}
On the other hand, if $\phi$ also derives the background expansion with Hubble rate $H$, one could define the corresponding Hubble slow-roll parameter as
\begin{align}
\epsilon_H\equiv-\frac{\dot{H}}{H^2}=\frac{\dot{\phi}^2}{2M_\mathrm{Pl}^2H^2},
\end{align}
where we use the Friedmann equation $M_\mathrm{Pl}^2\dot{H}=-\frac12\dot{\phi}^2$ in the last step. The e-folding number during given field excursion in the non-eternal inflationary regime is well-known,
\begin{align}\label{eq:dNdphi}
|\Delta N|&=\int_{t_i}^{t_f}H\mathrm{d}t=\int_{\phi_i}^{\phi_f}\frac{H}{\dot{\phi}}\mathrm{d}\phi=\int_{\phi_i}^{\phi_f}\frac{1}{\sqrt{2\epsilon_H}}\frac{|\mathrm{d}\phi|}{M_\mathrm{Pl}}\nonumber\\
&\equiv\left\langle\frac{1}{\sqrt{\epsilon_H}}\right\rangle\frac{|\Delta\phi|}{\sqrt{2}M_\mathrm{Pl}},
\end{align}
where in the second line we define the averaged value for the inverse square-root of Hubble slow-roll parameter. Therefore, Eq. \eqref{eq:NsDphi} becomes
\begin{align}\label{eq:NsN1}
N_s\gtrsim\mathrm{e}^{\frac23|\Delta N|/\langle\epsilon_H^{-1/2}\rangle}\equiv\mathrm{e}^{\frac23N_\epsilon},
\end{align}
where the abbreviation $N_\epsilon\equiv|\Delta N|/\langle\epsilon_H^{-1/2}\rangle$ is introduced for convenience. The physical meaning of bound \eqref{eq:NsN1} is illuminating: for given inflationary potential in an effective theory valid below $\Lambda_s$, the e-folding number is bounded from above by \eqref{eq:NsN1}, otherwise larger e-folding number leads to larger field excursion, and hence larger number of light particle states would be included to spoil the effectiveness of original theory unless lowering $\Lambda_s$ accordingly.

\subsection{A strong scalar weak gravity conjecture}\label{subsec:SSWGC}

However, there are some concerns \cite{Gonzalo:2019gjp}  on the scalar WGC, for example, it does not apply to all scalar fields (e.g. the massless mediator $\phi$ itself), and it does not simultaneously accomodate SDC for both $\phi\to\pm\infty$. Furthermore, it is not consistent with axion-like particles and fifth-force constraints, and it might be be in phenomenological tension \cite{Shirai:2019tgr} with the (refined) swampland dS conjecture \cite{Obied:2018sgi,Dvali:2018fqu,Andriot:2018wzk,Rudelius:2019cfh,Ooguri:2018wrx,Garg:2018reu,Andriot:2018mav}. As a result, a stronger version of scalar WGC was proposed in \cite{Gonzalo:2019gjp} for any canonically normalized real scalar $\phi$ coupled to gravity. The scalar potential $V(\phi)$ at any field value should meet
\begin{align}
2(V''')^2-V''V''''\geq\frac{(V'')^2}{M_\mathrm{Pl}^2},
\end{align}
which, after writing $m^2=V''$, becomes
\begin{align}\label{eq:SSWGC}
2(\partial_\phi m^2)^2-m^2(\partial_\phi^2m^2)\geq\frac{(m^2)^2}{M_\mathrm{Pl}^2}.
\end{align}
The above bound posses clear physical meaning \cite{Kusenko:2019kcu} that the first and second terms on the left-hand-side denote the attractive and repulsive scalar forces, respectively. Therefore, this strong version of scalar WGC simply states that the net scalar force should surpass the gravity force over the whole field space.

To see whether the bound \eqref{eq:NsN1} is still held in the strong version of scalar WGC, we could insert \eqref{eq:Ns} into \eqref{eq:SSWGC} and found
\begin{align}
9(\partial_\phi\ln N_s)^2+3\partial_\phi^2\ln N_s\gtrsim M_\mathrm{Pl}^{-2},
\end{align}
which could be directly solved as
\begin{align}\label{eq:NsN}
N_s\gtrsim \cosh^\frac13\frac{|\Delta\phi|}{M_\mathrm{Pl}}\gtrsim\mathrm{e}^{\frac{\sqrt{2}}{3}N_\epsilon}.
\end{align}
Apart from a $\mathcal{O}(1)$ factor difference in the exponent, this is the same as Eq. \eqref{eq:NsN1}, which could also be derived from applying SDC on \eqref{eq:Ns} in specific form of $m(\phi)$ with $m(\phi_0)\sim M_\mathrm{Pl}$. However, our derivation of \eqref{eq:NsN} is quite generic in the sense that it follows for any scalar field at any field value, and does not rely on the slow-roll approximation.

\section{Refined trans-Planckian censorship conjecture}\label{sec:TCC}

Armed with Eq. \eqref{eq:NsN}, a refined TCC could be targeted with following two separate arguments: 

\subsection{UV-IR hierarchy}\label{subsec:hierarchy}

The Hubble horizon as an IR cutoff scale should be bounded by the UV cutoff scale set by the species scale \cite{Brahma:2019vpl}, 
\begin{align}\label{eq:UVIR}
H<\Lambda_s\equiv\frac{M_\mathrm{Pl}}{\sqrt{N_s}}\lesssim M_\mathrm{Pl}\mathrm{e}^{-\frac{\sqrt{2}}{6}N_\epsilon},
\end{align}
with Eq. \eqref{eq:NsN} used in the last step, from which  a refined TCC could be directly read off,
\begin{align}\label{eq:TCC2}
\mathrm{e}^{\frac{\sqrt{2}}{6}N_\epsilon}\lesssim\frac{M_\mathrm{Pl}}{H}.
\end{align}
The original TCC \eqref{eq:TCC} serves as a stronger statement than \eqref{eq:TCC2} since $N_\epsilon\ll |\Delta N|$ for slow-roll inflation with $\epsilon_H\ll1$. Therefore, all those stingy examples that support original TCC also automatically hold for the refined TCC. The physical consequence of refined TCC could be understood in the following way: for given potential with expansion history $\langle\epsilon_H^{-1/2}\rangle$ fixed, the violation occurs for large enough $|\Delta N|$, which leads to large field excursion $|\Delta\phi|$. Hence  the species number $N_s$ would be so large that the UV scale set by species scale $\Lambda_s$ could be even smaller than the IR scale set by Hubble horizon $H$, which is not allowed by the hierarchy of scales.

\subsection{Entropy bounds}\label{subsec:entropy}

Consider a spherical region of size $R$ consisting of $N_s$ species number of particles as radiations at temperature $T$,  the total energy and entropy scale as
\begin{align}\label{eq:SR}
E\sim N_sR^3T^4, \quad S\sim N_sR^3T^3.
\end{align}
The absence of black hole formation imposes $R>R_\mathrm{BH}\sim E/M_\mathrm{Pl}^2\sim N_s R^3T^4/M_\mathrm{Pl}^2$, which leads to a maximum radius $R<M_\mathrm{Pl}/(\sqrt{N_s}T^2)$, or equivalently a maximum entropy $S\sim N_sR^3T^3<M_\mathrm{Pl}^3/(\sqrt{N_s}T^3)$. This maximum entropy should be bounded by the Bekenstein-Hawking entropy, $S_\mathrm{BH}\sim M_\mathrm{Pl}^2R^2<M_\mathrm{Pl}^4/(N_sT^4)$, as
\begin{align}\label{eq:SRSBH}
\frac{M_\mathrm{Pl}^3}{\sqrt{N_s}T^3}<\frac{M_\mathrm{Pl}^4}{N_sT^4},
\end{align}
which in turn gives rise to a maximum temperature $T<M_\mathrm{Pl}/\sqrt{N_s}$ \cite{Dvali:2007wp,Brustein:1999md,Brustein:2001di} larger than the Hubble temperature,
\begin{align}
\frac{H}{2\pi}<\frac{M_\mathrm{Pl}}{\sqrt{N_s}}\lesssim M_\mathrm{Pl}\mathrm{e}^{-\frac{\sqrt{2}}{6}N_\epsilon},
\end{align}
namely,
\begin{align}
\mathrm{e}^{\frac{\sqrt{2}}{6}N_\epsilon}\lesssim\frac{M_\mathrm{Pl}}{H}.
\end{align}
Similar bound could also be obtained if the radiation entropy is bounded by the Hubble entropy \cite{Veneziano:1999ty,Brustein:2007hd},
\begin{align}\label{eq:SRSH}
N_sR^3T^3<M_\mathrm{Pl}^2R^3H.
\end{align}
Hence there is also a maximum temperature, $T<(M_\mathrm{Pl}^2H/N_s)^{1/3}$, which should be larger than the Hubble temperature,
\begin{align}
H^3<\frac{M_\mathrm{Pl}^2H}{\sqrt{N_s}}\lesssim M_\mathrm{Pl}^2H\mathrm{e}^{-\frac{\sqrt{2}}{6}N_\epsilon},
\end{align}
namely,
\begin{align}
\mathrm{e}^{\frac{\sqrt{2}}{12}N_\epsilon}\lesssim\frac{M_\mathrm{Pl}}{H}.
\end{align}
Note that similar arguments of using entropy bounds are adopted in \cite{Kehagias:2019iem} to elaborate SDC.

The above arguments implicitly assume that all $N_s$ species number of particles have their masses spectrum below the temperature. However, it is fairly possible that some of the particle masses are heavier than the temperature so that they have contribution to the total energy $E\sim k_BN_sT$ ($k_B\equiv1$ hereafter) but no contribution to the total entropy, which should be bounded by the Bekenstein entropy \cite{Bekenstein:1972tm,Bekenstein:1973ur,Bekenstein:1974ax} for any weakly gravitating matter fields,
\begin{align}\label{eq:SB}
S_\mathrm{matter}\leq2\pi ER\sim2\pi N_sTR.
\end{align}
If the matter inside the sphere is gravitationally stable so that black hole does not form, the Bekenstein entropy should be further bounded by the Bekenstein-Hawking entropy,
\begin{align}\label{eq:SBSBH}
2\pi N_sTR\leq M_\mathrm{Pl}^2R^2.
\end{align}
Therefore, there is a maximum temperature $T<M_\mathrm{Pl}^2R/(2\pi N_s)$, which should be larger than the Hubble temperature, 
\begin{align}
\frac{H}{2\pi}<\frac{M_\mathrm{Pl}^2R}{2\pi N_s}\lesssim\frac{M_\mathrm{Pl}^2}{2\pi H}\mathrm{e}^{-\frac{\sqrt{2}}{3}N_\epsilon},
\end{align}
namely,
\begin{align}
\mathrm{e}^{\frac{\sqrt{2}}{6}N_\epsilon}\lesssim\frac{M_\mathrm{Pl}}{H}.
\end{align}
The same bound could also be obtained by bounding the Bekenstein entropy with Hubble entropy , 
\begin{align}\label{eq:SBSH}
2\pi N_sTR\leq M_\mathrm{Pl}^2R^3H.
\end{align}
Hence there is also a maximum temperature $T<M_\mathrm{Pl}^2HR^2/(2\pi N_s)$, which should be larger than the Hubble temperature, 
\begin{align}
\frac{H}{2\pi}<\frac{M_\mathrm{Pl}^2HR^2}{2\pi N_s}\lesssim\frac{M_\mathrm{Pl}^2}{2\pi H}\mathrm{e}^{-\frac{\sqrt{2}}{3}N_\epsilon},
\end{align}
namely, 
\begin{align}
\mathrm{e}^{\frac{\sqrt{2}}{6}N_\epsilon}\lesssim\frac{M_\mathrm{Pl}}{H}.
\end{align}
Note that the entropy bound alone is not sufficient to guarantee the refined TCC, since the UV-IR hierarchies $H<T$ and  $R<H^{-1}$ are still needed for the relevant derivation of inequalities. 

In summary, with Eq. \eqref{eq:NsN}, the refined TCC is always implied by entropy bounds either from radiation entropy \eqref{eq:SR} or Bekenstein entropy \eqref{eq:SB}, which are bounded by either Bekenstein-Hawking entropy \eqref{eq:SRSBH}\eqref{eq:SBSBH} or Hubble entropy \eqref{eq:SRSH}\eqref{eq:SBSH}. 

\section{Eternal inflation}\label{sec:eternal}

There are three types of eternal inflation frequently discussed in the literatures: old eternal inflation (a trapped dS vacuum) \cite{Guth:1980zm,Guth:1982pn}, stochastic eternal inflation \cite{Vilenkin:1983xq,Starobinsky:1986fx,Linde:1986fc,Aryal:1987vn}  (a hilltop potential \cite{Barenboim:2016mmw}), and topological eternal inflation \cite{Vilenkin:1994pv,Linde:1994hy,Linde:1994wt} (an eternally inflating topological defect from a hilltop potential \cite{Boubekeur:2005zm}). Whether the eternal inflation is in the swampland is an important question currently under debate \cite{Matsui:2018bsy,Dimopoulos:2018upl,Kinney:2018kew,Brahma:2019iyy,Wang:2019eym,Rudelius:2019cfh,Blanco-Pillado:2019tdf,Lin:2019fdk}. The old eternal inflation is generically inconsistent with  the original SdSC \cite{Matsui:2018bsy,Dimopoulos:2018upl}, but the stochastic eternal inflation was argued to be marginally consistent with the refined SdSC \cite{Kinney:2018kew}, which, however, was questioned with perturbative \cite{Brahma:2019iyy} and entropic \cite{Wang:2019eym} arguments. Nevertheless, \cite{Rudelius:2019cfh} speculated a possibility that some variants of SdSC should be regarded as approximate consequences of the absence of eternal inflation. Furthermore, some topological eternal inflation models could be constructed to evade the swampland criteria for an eternally inflating bubble wall \cite{Blanco-Pillado:2019tdf} and domain wall on the brane \cite{Lin:2019fdk}. 

\begin{figure}
\centering
\includegraphics[width=0.47\textwidth]{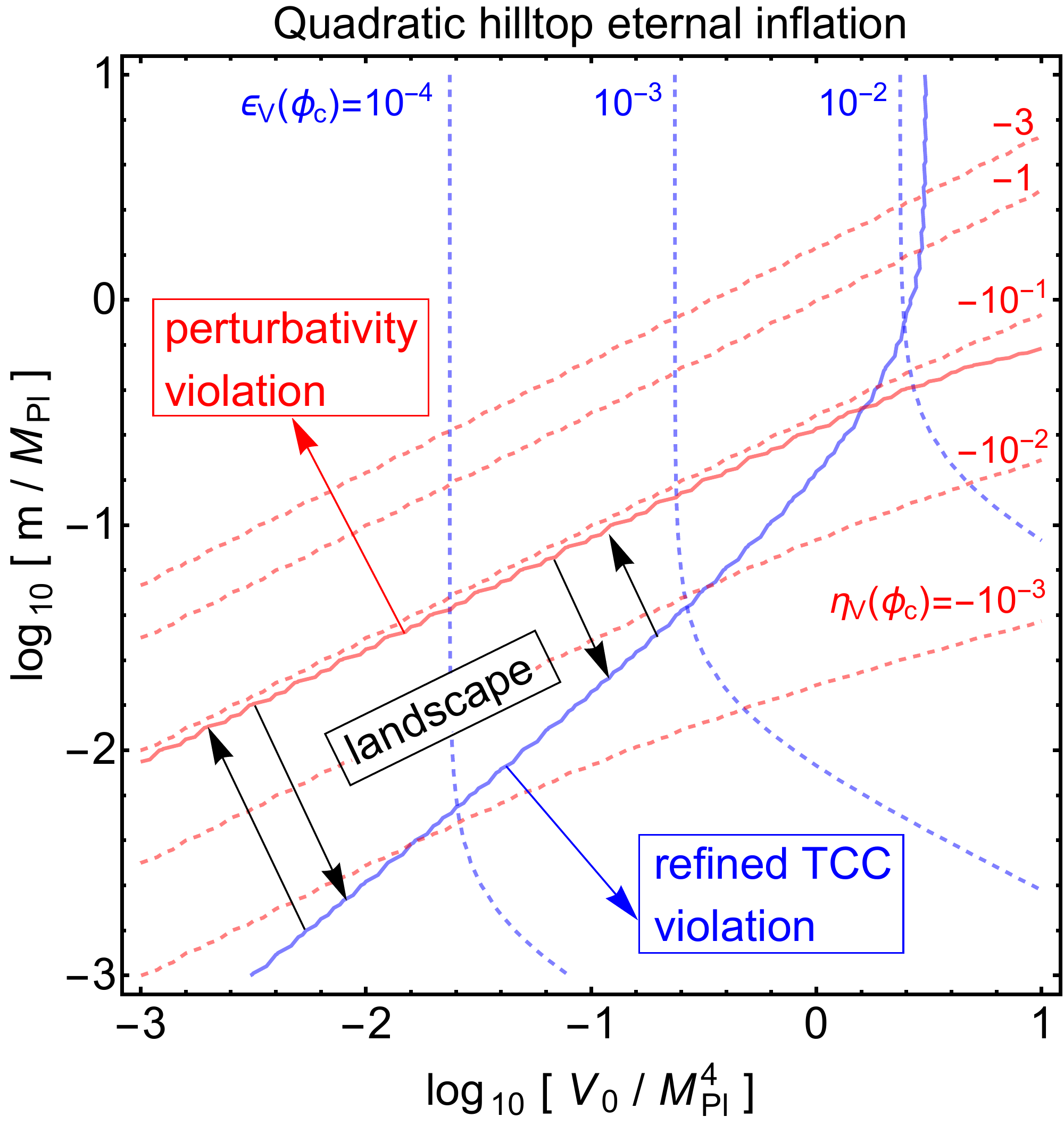}\\
\caption{The parameter space for a quadratic hilltop eternal inflation out of swampland by our refined TCC enclosed by the red and blue solid lines. The slow-roll approximation is also checked with the blue and red dashed lines.}\label{fig:hilltop}
\end{figure}

Our refined TCC \eqref{eq:TCC2} is elaborated from the arguments of Eq. \eqref{eq:NsDphi}, \eqref{eq:dNdphi} and \eqref{eq:UVIR}. Note that Eq. \eqref{eq:dNdphi} is restricted to the non-eternal inflationary regime where the field excursion is dominated by the classical rolling. For eternal inflation, the elapsed e-folding number could be arbitrarily large as long as the field excursion is limited in the range where the quantum fluctuations dominate over classical rolling. In this case, one could use \eqref{eq:NsDphi} and  \eqref{eq:UVIR} directly without \eqref{eq:dNdphi} to render the general version of our refined TCC, 
\begin{align}\label{eq:eternal0}
\mathrm{e}^{\frac{\sqrt{2}}{6}\frac{|\Delta\phi|}{M_\mathrm{Pl}}}\lesssim\frac{M_\mathrm{Pl}}{H},
\end{align}
which shares a similar form derived in \cite{Scalisi:2018eaz} from SDC on inflation. To see whether the stochastic eternal inflation could survive above criterion, we use a quadratic hilltop potential
\begin{align}\label{eq:hilltop}
V(\phi)=V_0-\frac12m^2\phi^2, \quad m^2>0,
\end{align}
where the eternal inflation is allowed in the small field region $|\Delta\phi|<\phi_c$ when the averaged quantum random walks dominate over classical rolling during each Hubble time, 
$H/(2\pi)\equiv\langle\delta\phi\rangle_Q>\delta\phi_C\equiv|\dot{\phi}|/H$.
$\phi_c$ is therefore solved from
\begin{align}\label{eq:phic}
\mathcal{P}_\zeta(\phi_c)\equiv\left(\frac{H^2(\phi_c)}{2\pi\dot{\phi}_c}\right)^2\approx
\frac{V(\phi_c)}{24\pi^2M_\mathrm{Pl}^4\epsilon_V(\phi_c)}=1.
\end{align}
Our refined TCC \eqref{eq:eternal0} is ensured provided that
\begin{align}\label{eq:eternalc}
\mathrm{e}^{\frac{\sqrt{2}}{6}\frac{\phi_c}{M_\mathrm{Pl}}}<\frac{M_\mathrm{Pl}}{H(\phi_c)},
\end{align}
which could be solved numerically as shown in Fig. \ref{fig:hilltop} with blue solid line.
Furthermore, \eqref{eq:eternalc} automatically guarantees the slow-roll condition for $\epsilon_V$ (blue dashed lines),
\begin{align}
\epsilon_V(\phi_c)<\frac{1}{8\pi^2}\mathrm{e}^{-\frac{\sqrt{2}}{3}\frac{\phi_c}{M_\mathrm{Pl}}}<1.
\end{align}
The slow-roll condition $|\eta_V(\phi_c)|\ll1$ for $\eta_V(\phi_c)$ (red dashed lines) also automatically guarantees $\eta_V>-3$ \cite{Rudelius:2019cfh} so that the exponentially suppressed probability to stay within $|\Delta\phi|<\phi_c$ beats the exponential expansion of the universe. On the other hand, the perturbativity argument \cite{Brahma:2019iyy} further requires
\begin{align}
\mathcal{P}_\zeta<\frac{1}{16[3(\eta_V-\epsilon_V)-\epsilon_V(\eta_V-\epsilon_V)-\epsilon_V\eta_V]^2},
\end{align}
which is shown in Fig. \ref{fig:hilltop} with red solid line. Therefore, there is a parameter space for the quadratic hilltop eternal inflation out of swampland by our refined TCC \eqref{eq:eternal0}.

\section{Conclusion}\label{sec:con}

In this paper, we proposed a refined TCC as a consequence of a strong version of scalar WGC combined with either UV-IR hierarchy or some entropy bounds,
\begin{align}\label{eq:eternal}
\mathrm{e}^{\frac{|\Delta\phi|}{M_\mathrm{Pl}}/\mathcal{O}(1)}\lesssim\mathcal{O}(1)\frac{M_\mathrm{Pl}}{H},
\end{align}
which is consistent with the hilltop eternal inflation. For non-eternal inflation, it further reduces to
\begin{align}\label{eq:TCC3}
\mathrm{e}^{N_\epsilon/\mathcal{O}(1)}\lesssim\mathcal{O}(1)\frac{M_\mathrm{Pl}}{H_f}.
\end{align}
where the $\mathcal{O}(1)$ factor on the left-hand-side is associated to certain combination of spacetime dimension, which could be easily worked out from previous arguments, and the $\mathcal{O}(1)$ factor on the right-hand-side is already mentioned in a footnote in \cite{Bedroya:2019snp}, which could be the sound speed in the generalized TCC in \cite{Lin:2019pmj}. The presence of an averaged Hubble slow-roll parameter in the exponent  $N_\epsilon\equiv N/\langle\epsilon_H^{-1/2}\rangle$ would prolong the dS lifetime by
\begin{align}
\Delta t_\mathrm{dS}<\frac{1}{H_f}\int_{t_i}^{t_f}H\mathrm{d}t=\frac{1}{H_f}\ln\frac{a_f}{a_i}<\frac{\langle\epsilon_H^{-1/2}\rangle}{H_f}\ln\frac{M_\mathrm{Pl}}{H_f},
\end{align} 
which resembles the form of the entropic quasi-de Sitter instability time \cite{Seo:2019wsh} (such a logarithmic correction to the Hubble time also appears in the scrambling time \cite{Sekino:2008he,Bedroya:2019snp}). Now the constraint on the Hubble scale could be greatly relaxed, for instance, $N_\epsilon\equiv N/\langle\epsilon_H^{-1/2}\rangle\approx N/10=6$ for an illustrative $\epsilon_H\approx0.01$ and e-folding number $N=60$, to 
\begin{align}
H\lesssim\mathrm{e}^{-N_\epsilon}M_\mathrm{Pl}\approx6\times10^{15}\,\mathrm{GeV}.
\end{align} 
The same relaxation could also be achieved even for a near critical expansion with $\epsilon_H\approx1$ due to the presence of a $\mathcal{O}(1)$ factor on the left-hand-side of \eqref{eq:TCC3}, for example, as shown in \eqref{eq:TCC2},
\begin{align}
H\lesssim\mathrm{e}^{-\frac{\sqrt{2}}{6}N}M_\mathrm{Pl}\approx2\times10^{12}\,\mathrm{GeV},
\end{align} 
for the same illustrative $N=60$. Therefore, our refined TCC resolves the fine-tuning problem of inflationary initial conditions without invoking any exotic model-building. Furthermore, since our refined TCC is a weaker version of original TCC, it would be automatically supported by those stringy examples that also hold for the original TCC. 

\section{Discussions}\label{sec:diss}

Several comments are given in order  below  regarding several concerns on our refined TCC.

Firstly, although initially motivated to refine the original TCC constraint on the inflationary Hubble scale, we haven't directly addressed the trans-Planckian problem yet. Similar to the implication from the entropic quasi-de Sitter instability time \cite{Seo:2019wsh}, the physical length of those modes at the beginning of inflation that freeze out the Hubble horizon at the end of inflation is
\begin{align}
k_\mathrm{max}^{-1}=\frac{a_i}{a_fH_f}=\frac{1}{e^{N_\mathrm{inf}}H_f},
\end{align}
which, after replacing the total inflationary e-folding number $N_\mathrm{inf}$ with saturation of our TCC bound for non-eternal inflation \eqref{eq:TCC3},
\begin{align}
e^{N_\mathrm{inf}^\mathrm{max}}\simeq\left(\mathcal{O}(1)\frac{M_\mathrm{Pl}}{H_f}\right)^{\mathcal{O}(1)\langle\epsilon_H^{-1/2}\rangle},
\end{align}
allows for trans-Planckian modes,
\begin{align}
k_\mathrm{max}\simeq H_f\left(\mathcal{O}(1)\frac{M_\mathrm{Pl}}{H_f}\right)^{\mathcal{O}(1)\langle\epsilon_H^{-1/2}\rangle}>\mathcal{O}(1)M_\mathrm{Pl}.
\end{align}
This is not surprising since the original TCC that exactly forbids the exit of trans-Planckian modes is refined (weaken) in our formulation. 
Furthermore, arguments that put forward recently in \cite{Berera:2020dvn} against the Hubble horizon being a scale of singular significance of trans-Planckian problem have shown that classicalization does not necessarily require a trans-Planckian mode to cross the Hubble horizon at all. Therefore, we would not consider the allowance of trans-Planckian modes as a serious problem for our refined TCC. Nevertheless, it also weakens the connection to original TCC that is closely attached to the trans-Planckian problem. As a result, our refined TCC might be better regarded as another weaker version among other swampland conjectures.

Secondly, as one could elaborate from the derivation of our refined TCC, relation \eqref{eq:Ns} (or more precisely $N_s\sim\Lambda_s/m$) is a crucial assumption. In fact, $N_s\sim\Lambda_s/m$ follows closely to SDC that, on large distance in field space, $|\Delta\phi|>M_\mathrm{Pl}$, a tower of light states with mass and number
\begin{align}
m(\phi)&\sim m(\phi_0)e^{-\alpha|\Delta\phi|/M_\mathrm{Pl}},\\
N_s(\phi)&\sim e^{\alpha|\Delta\phi|/M_\mathrm{Pl}},
\end{align}
appear descending from the UV if $m(\phi_0)\sim\Lambda_s$ is identified. Although this relation has been widely used in the literatures (see, for example, \cite{Palti:2019pca} around Eq. (5.17) and references therein), it certainly should merits further rigorous elaboration in future.

Thirdly, the naive combination of scalar WGC with either UV-IR hierarchy or some entropy bounds leads to \eqref{eq:eternal}, which is nothing but a weaker version of SDC,
\begin{align}
\frac{|\Delta\phi|}{M_\mathrm{Pl}}\lesssim\mathcal{O}(1)\ln\left(\mathcal{O}(1)\frac{M_\mathrm{Pl}}{H}\right),
\end{align}
with extra logarithmic correction. In fact, this relaxed form of SDC could easily meet the current observational constraint on inflation by expressing the field excursion with Lyth bound and Hubble scale with $A_s=2.1\times10^{-9}$,
\begin{align}
|\Delta N|\sqrt{\frac{r}{8}}\lesssim\mathcal{O}(1)\ln\left(\frac{\sqrt{2}\mathcal{O}(1)}{\pi\sqrt{rA_s}}\right),
\end{align}
which constrains the tensor-to-scalar ratio as
\begin{align}
r\lesssim\frac{8\mathcal{O}(1)^2}{\Delta N^2}W\left(\frac{\mathcal{O}(1)|\Delta N|}{2\pi\mathcal{O}(1)\sqrt{A_s}}\right)^2\sim\mathcal{O}(0.1-1)
\end{align}
for typical $|\Delta N|=60$. To further ensure the validity of effective theory of inflation,  the mass scale of infinite tower of light states due to the field excursion $|\Delta\phi|=|\phi_f-\phi_i|$ should be larger than the inflationary  energy scale $V^{1/4}\simeq M_\mathrm{Pl}^{1/2}H^{1/2}$,
\begin{align}
m(\phi_i)e^{-\alpha|\Delta\phi|/M_\mathrm{Pl}}
&\gtrsim m(\phi_i)\left(\mathcal{O}(1)\frac{M_\mathrm{Pl}}{H}\right)^{-\alpha\mathcal{O}(1)}\nonumber\\
&>M_\mathrm{Pl}^{1/2}H^{1/2},
\end{align}
which constrains the mass scale of infinite tower of states before field excursion by
\begin{align}
m(\phi_i)\gtrsim\mathcal{O}(1)^{\alpha\mathcal{O}(1)}\left(\frac{M_\mathrm{Pl}}{H}\right)^{\alpha\mathcal{O}(1)-\frac12}M_\mathrm{Pl}.
\end{align}
The above procedures are similar to the normal constraint on inflation from original SDC. The upshot here is that, the relaxed form of SDC does not necessarily violate the ordinary SDC or observational bounds, nor does it forbid the super-Planckian field excursion as long as $m(\phi_f)>V^{1/4}$.

Finally, our refined TCC \eqref{eq:eternal} seems to allow for stable dS vacua when field excursion $\Delta\phi$ is vanished, and it is too weak to constrain the effective field theory of quantum gravity. On the other hand, this is not surprising since our starting point is (strong) scalar WGC, which is weakest among all the other swampland conjectures. Now that it is rather difficult to either verify or falsify other stronger swampland conjectures, our refined TCC could still find its position by serving as the bottom-line constraint on the effective theory of inflation from swampland program.

\begin{acknowledgments}
We thank Suddhasattwa Brahma, Mark Hertzberg, Ken Olum, Eran Palti, Fabrizio Rompineve, Shan-Ming Ruan, Alexander Vilenkin, and Masaki Yamada for helpful discussions and correspondences.
R.G.C. was supported by the National Natural Science Foundation of China Grants No. 11435006, No. 11647601, No. 11690022, No. 11821505 and No. 11851302, and by the Strategic Priority Research Program of CAS Grant No. XDB23030100, and by the Key Research Program of Frontier Sciences of CAS. SJW is supported by the postdoctoral scholarship of Tufts University from NSF.
\end{acknowledgments}


\bibliographystyle{utphys}
\bibliography{ref}

\end{document}